# Influence of resonances on the $^{11}$B(n,γ)$^{12}$B capture reaction rate. III. Capture to the 2$^{nd}$, 3$^{rd}$ and 4$^{th}$ excited state of $^{12}$B.

## Dubovichenko S.B.[1,2,*], Burkova N.A.[2], Dzhazairov-Kakhramanov A.V.[1,*]

[1]Fesenkov Astrophysical Institute "NCSRT" ASA MDASI RK, 050020, Almaty, Kazakhstan
[2]al-Farabi Kazakh National University, 050040, Almaty, Kazakhstan

**Abstract:** Within the framework of the modified potential cluster model with a classification of orbital states according to Young diagrams, the possibility of data description of the radiative neutron capture on $^{11}$B to the second excited state of $^{12}$B at 1.67365 MeV (2$^-$) and prediction absentee of experimental data for the total cross sections to the third and fourth excited state of $^{12}$B at 2.6208 MeV (1$^-$) and 2.723 MeV (0$^+$) for reaction energy of 10 meV (1 meV = 10$^{-3}$ eV) to 7 MeV was considered. The reaction rate in the temperature range of 0.01 to 10.0 $T_9$ is calculated on the basis of obtained cross sections, which take into account resonances up to 5 MeV. It is shown, that low-lying resonances exercise a significant influence to the capture reaction rate. The approximation of the calculation reaction rate is carried out by the simple analytic formula.

**Keywords:** nuclear astrophysics, light atomic nuclei, low and astrophysical energies, elastic scattering, $n^{11}$B system, $^{12}$B nucleus, first excited state, potential cluster model, radiative capture, total cross sections, thermonuclear reactions, forbidden states, Young diagrams.

**PACS No.:** 21.60.Gx, 25.20.-x, 25.40.Lw

## 1. Introduction

Earlier, the neutron capture process on $^{11}$B to the ground state (GS) and the first excited state (1$^{st}$ ES) of $^{12}$B was considered by us within the framework of the modified potential cluster model (MPCM) in [1,2]. On the basis of the MPCM, it is possible to correctly describe available experimental data for capture to the GS and to obtain quite reasonable results for the consideration capture to the 1$^{st}$ ES, for which the experimental data are absent. Continuing studying the processes of the radiative capture on the basis of the MPCM [3,4], let us consider capture reaction $n^{11}$B → γ$^{12}$B to the second, third and fourth excited states (2$^{nd}$, 3$^{rd}$, 4$^{th}$ ESs) of $^{12}$B at $J^\pi$ = 2$^-$, 1$^-$ and 0$^+$ with excitation energy of 1.67365, 2.6208 and 2.723 MeV and will take into account resonance levels (RLs) in the continuous spectrum up to 5 MeV above the threshold of the $n^{11}$B channel.

In the present calculations we will use new data on spectra of $^{12}$B from [5], comparably with our previous papers [6,7], where the capture to the ground state (GS) and few excited states (ESs) were considered, wherein results of earlier review [8] were used. In our previous work [6] old data on asymptotic constants (ACs) of the $n^{11}$B system were used, which did not take into account all possible transitions to all ESs and all resonances at low energies. Although, we take into account more modern data on AC in our work [7], the capture only to the GS without resonances was considered, as it was done by us in the latest work [1].

Note at once that at energies above 100 keV only experimental data [9] are known to us. Other measurements in data bases, for example, [10] or [11] are absent. It is impossible for us to find earlier theoretical calculations for this reaction taken into account all possible resonances and captures to all ESs. At the same time our calculations more than 30 reactions of the radiative capture [3,4] and [12] (and refs for it) show the efficiency of the used by us MPCM for such processes. Therefore, there is a cause to think that the model in this case too will allow to obtain reasonable results for considered reaction.

---
[*] Corresponding authors: albert-j@yandex.kz, dubovichenko@gmail.com

In the present work, apparently, the capture to the pointed above three excited states of $^{12}$B is considered taking into account the resonances in the scattering process of initial particles of the input channel. The considered here reaction $n^{11}\text{B} \to \gamma^{12}\text{B}$ was studied earlier in [13], where for the construction of the $n^{11}$B interaction potentials the thermal neutron capture cross section $\sigma_{th}$ equals 5.5(3.3) mb was used [14]. At the same time, more modern data from [15], where the value $\sigma_{th} = 9.09(10)$ mb is given, were used in our work [1].

## 2. Calculation methods and state structure

Expressions for total cross sections of the radiative capture $\sigma(NJ,J_f)$ for *EJ* and *MJ* transitions in the potential cluster model are given, for example, in [16] or [3,4] and for *MJ* transitions in [1]. The values from data base [17] and work [18] were used for cluster magnetic moments, that is, $\mu_n = -1.91304272\mu_0$ and $\mu(^{11}\text{B}) = 2.6887\mu_0$. The next mass values of particles were used in calculations: $m_n = 1.00866491597$ amu [12], $m(^{11}\text{B}) = 11.0093052$ amu [14], and constant $\hbar^2/m_0$ is equal to 41.4686 MeV·fm$^2$, where $m_0$ is atomic mass unit (amu). Calculation methods in the frame of the MPCM of other values, for example, root-mean-square mass and charge radii or binding energy, which are considered further, are given, for example, in [3,4].

The used intercluster potentials do not contain ambiguities and, as it was shown in [3,4], allow one to correctly describe the total cross sections of many processes of radiative capture. The potentials of the bound states (BSs), namely bound ESs, should correctly describe the known values of the AC, which is associated with the asymptotic normalization coefficient (ANC) of the $A_{NC}$ usually extracted from the experiment as follows [20] (in our notations)

$$A_{NC}^2 = S_f \cdot C^2, \qquad (1)$$

where $S_f$ is the spectroscopic factor of the considered channel and *C* is the dimensional AC, expressed in fm$^{-1/2}$ and determined from the relation

$$\chi_L(r) = C \cdot W_{-\eta L+1/2}(2k_0 r). \qquad (2)$$

It is related to the dimensionless AC $C_w$ [21], used by us, as follows: $C = \sqrt{2k_0} C_w$, and the dimensionless constant $C_w$ is defined by expression [21]

$$\chi_L(r) = \sqrt{2k_0} \cdot C_w \cdot W_{-\eta L+1/2}(2k_0 r). \qquad (3)$$

Here $\chi_L(r)$ is the numerical wave function of the bound state obtained from the solution of the radial Schrödinger equation and normalized to unit. $W_{-\eta L+1/2}$ is the Whittaker function of the bound state, which determines the asymptotic behavior of the wave function (WF) and is a solution of the same equation without nuclear potential, that is, at large distances $r = R$; $k_0$ is the wave number caused by the channel binding energy; η is the Coulomb parameter, which is equal to zero in this case, and $L$ is the orbital momentum of this BS.

Let us assume furthermore that for $^{11}$B (spin and isospin of $^{11}$B are equal to $J^π,T = 3/2^-,1/2$ [5]) it is possible to take Young diagram in the form {443}, therefore for the $n^{11}$B system we have {1} × {443} → {543} + {444} + {4431} [22]. First of the obtained diagrams compatible with orbital momenta $L = 1,2,3,4$ and is forbidden (FS), because it cannot be five nucleons in the *s*-shell [23], the second diagram {444} apparently corresponds to the allowed state (AS) compatible with orbital momenta $L = 0,2,4$, and the third {4431}, also allowed, compatible with $L = 1,2,3$ [23].

Thus, limiting only by the low partial waves with orbital momenta $L = 0,1,2,3$, it is possible to say that there is only AS for diagram {444} in the $S$ wave potential of the $n^{11}$B system. There are forbidden {543} and allowed {4431} states in $P$ waves. In particular, the GS of $^{12}$B with momentum $J^π = 1^+$ corresponds to the $P_1$ wave with {4431}, lying at binding energy of the $n^{11}$B system of -3.370 MeV [5]. FS with the diagram {543} and AS at {4431}+{444} correspond to the $D$ waves. FS with {543} and AS with {4431} correspond to the $F$ waves. These ASs for scattering potentials can lie in the continuous spectrum and are unbounded. The AS can correspond to the 1$^{st}$ ES of $^{12}$B with momenta $J^π = 2^+$, lying at the binding energy of the $n^{11}$B system of -2.4169 MeV [5]

## 3. Possible transitions in the $n^{11}$B system to the 2$^{nd}$, 3$^{rd}$ and 4$^{th}$ ESs of $^{12}$B

Let us consider now available excited states in $^{12}$B nucleus, but bound in the $n^{11}$B channel [1,2].

1. There is first excited state (1$^{st}$ ES), but not bound in this channel, with momentum of $J^π = 2^+$, which can be matched to the $^{3+5}P_2$ wave with the bound FS, at the excitation energy of 0.95314(60) MeV or -2.41686(60) MeV [5] relatively to the threshold of the $n^{11}$B channel.

2. Second excited state (2$^{nd}$ ES) at excitation energy of 1.67365(60) MeV [8] relatively to the GS or -1.69635(60) MeV relatively to the threshold of the $n^{11}$B channel has $J^π = 2^-$, and it can be matched to the $^5S_2$ wave without FS. In this case, also $^{3+5}D_2$ wave with the FS is possible.

3. Third excited state (3$^{rd}$ ES) at excitation energy of 2.6208(12) MeV [5] or -0.7492(12) MeV relatively to the threshold of the $n^{11}$B channel has $J^π = 1^-$, and it can be matched to the triplet $^3S_1$ wave without forbidden BS. In this case, also $^{3+5}D_1$ wave with the FS is possible.

4. Fourth excited state (4$^{th}$ ES) at excitation energy of 2.723(11) MeV [4] or -0.647(11) MeV [5] relatively to the threshold of the $n^{11}$B channel has $J^π = 0^+$, and it can be matched to the triplet $^3P_0$ wave with the bound FS.

Furthermore, the potentials of these ESs [1,2] listed in Tables 1, 2, 3 were constructed, which are using for description of the scattering processes in these partial waves that do not include resonances.

Besides excited states there are few resonance states (RSs), that is, states at positive energies relatively to the threshold of the $n^{11}$B channel [1,2].

1. First resonance state (1$^{st}$ RS) of $^{12}$B in the $n^{11}$B channel is at the excitation energy

of 3.3891(16) MeV or at neutron energy $E_n$ = 20.8(5) keV, it has the width of 1.4 keV in a center of mass (c.m.) and momentum of $J^\pi = 3^-$ (see Table 12.10 in [5]) – it can be matched to the $^{3+5}D_3$ scattering wave with forbidden bound state. The $E2$ transition to the 2$^{nd}$ ES $^5S_2$ of the form $^5D_3 \to {}^5S_2$ is possible; besides the $E2$ transition to the 3$^{rd}$ ES $^3S_1$ is possible too.

2. Second resonance state (2$^{nd}$ RS) lies at an energy of $E_n$ = 430(10) keV, its width in c.m. equals 37(5) keV and momentum of $J^\pi = 2^+$ [5]. Therefore, it can be matched to the $^{3+5}P_2$ scattering wave with forbidden bound state. The $E1$ transition to the 2$^{nd}$ ES $^5S_2$ of the form $^5P_2 \to {}^5S_2$, and also the $E1$ transition to the 3$^{rd}$ ES $^5S_2$ of the form $^3P_2 \to {}^3S_1$ and $E2$ transition to the 4$^{th}$ ES are possible.

3. Third resonance state (3$^{rd}$ RS) lies at an energy of $E_n$ = 1027(11) keV, its width in c.m. equals 9(4) keV and momentum of $J^\pi = 1^-$ [5] – it can be matched to the $^3S_1$ state without FS or $^{3+5}D_1$ scattering wave with the bound FS. In the last case, the $E2$ transition to the 2$^{nd}$ ES $^5S_2$ of the form $^5D_1 \to {}^5S_2$, the $E2$ transition to the 3$^{rd}$ ES $^5S_2$ of the form $^3D_1 \to {}^3S_1$ and to the 4$^{th}$ ES $^3P_0$ of the form $^3D_1 \to {}^3P_0$ are possible.

4. Eighth resonance state (8$^{th}$ RS) is at energy of $E_n$ = 2.580(20) MeV, with the width of 55(20) keV in c.m. and momentum of $J^\pi = 3^-$ can be matched to the $^{3+5}D_3$ scattering wave (see Table 12.10 [5]) and here, similar to the 1$^{st}$ RS, it is possible the $E2$ transition to the 2$^{nd}$ ES $^5S_2$ and to the 3$^{rd}$ ES $^3S_1$.

5. Furthermore, there is the resonance state (13$^{th}$ RS) with neutron energy of 4.70 MeV, with the width of 45 keV in c.m. and momentum of $J^\pi = 2^-$. It can be matched to the $^5S_2$ wave without FS or to the $^{3+5}D_2$ wave with the FS, which allows $E2$ transition to the 2$^{nd}$ ES $^5S_2$ of the form $^5D_1 \to {}^5S_2$, and to the 3$^{rd}$ ES $^3S_1$ of the form $^3D_2 \to {}^3S_1$.

6. The resonance state (14$^{th}$ RS) at the energy of 4.80 MeV, with the width of 90 keV in c.m. has momentum of $J^\pi = 1^-$. It can be matched to the $^3S_1$ wave without FS or to the $^{3+5}D_1$ wave with FS, which similar to the 3$^{rd}$ RS allows the $E2$ transition to the 2$^{nd}$ ES of the form $^5D_1 \to {}^5S_2$, to the 3$^{rd}$ ES $^3S_1$ and the $E1$ transition to the 4$^{th}$ ES $^3P_0$.

7. Higher states are studied till not so closely [5], and we will not consider them. Consequently, it is possible to consider influence of six resonances at energies up to $E_n$ = 5 MeV. Rests of resonances have either unknown width or momentum, therefore, it is impossible to construct unambiguous potentials, as it was done for other resonances. Besides, some resonances, as it was said in [1], are possible to be described only in assumption of the $F$ waves [1] and will not be considered. Here, all list of 14 resonances at excitation energies up to 5 MeV is analyzed in our works [1,2].

As was shown above, RS Nos. 3 and 6 by their momentum can coincide with the 3$^{rd}$ ES and No. 5 with the 2$^{nd}$ ES. However, it is not possible to construct $S$ potentials, which have bound AS, coincided with one of the ES and having the resonance at the observed excitation energy. Besides, these resonances cannot be described if to match them those $S$ potentials [1]. Therefore, the resonance potentials will be constructed so that they correspond to $D$ waves with FS and have the resonance at necessary energy with necessary width, and potentials for the bound ESs are corresponded to the $S$ waves.

Thus, as a result of analysis these ESs and RSs, furthermore we will consider transitions, which are listed in Tables 1, 2 and 3. Nonresonance transition No. 1 from Table 1, marked by italic, will not be considered so as the same $^5S_2$ potential in the continuous and discrete spectrum is used for it, and this leads to zero matrix elements for $M1$ transitions. It is given in the table in order to obtain potential parameters of this ES. Parameters of the $^3S_1$ potential of the 3$^{rd}$ ES, which is used for nonresonance scattering, are given in Table 2 under No. 1.

Table 1. The list of possible transitions from the $\{^{(2S+1)}L_J\}_i$ state to the different WF components $\{^{(2S+1)}L_J\}_f$ 2nd ES $^5S_2$ of $^{12}$B at the neutron capture on $^{11}$B and Gaussian parameters for initial scattering states.

| No. | $\{^{(2S+1)}L_J\}_i$ | Transition | $\{^{(2S+1)}L_J\}_f$ | $P^2$ | $V_0$, MeV | $\alpha$, fm$^{-2}$ | $E_r$, keV | $\Gamma_r$, keV |
|---|---|---|---|---|---|---|---|---|
| 1 | $^5S_2$ nonresonance scattering wave. | M1 | $^5S_2$ | --- | 6.70125 | 0.03 | --- | --- |
| 2 | $^5P_1$ nonresonance scattering wave – the GS potential was used | E1 | $^5S_2$ | 3 | 194.68751 | 0.22 | --- | --- |
| 3 | $^5P_2$ resonance at 430 keV – No.2. | E1 | $^5S_2$ | 5 | 11806.017 | 15.0 | 430 [430(10)] | 37 [37(5)] |
| 4 | $^5D_1$ resonance at 1.027 MeV – No.3. | E2 | $^5S_2$ | 3 | 1611.95103 | 1.25 | 1027 [1027(11)] | 9 [9(4)] |
| 5 | $^5D_1$ resonance at 4.8 MeV – No.6. | E2 | $^5S_2$ | 3 | 4502.245 | 3.5 | 4.80 [4.8] | 86 [90] |
| 6 | $^5D_2$ resonance at 4.7 MeV – No.5. | E2 | $^5S_2$ | 5 | 6444.382 | 5.0 | 4.70 [4.7] | 49 [45] |
| 7 | $^5D_3$ resonance at 20.8 keV – No.1. | E2 | $^5S_2$ | 7 | 16.0578 | 0.0125 | 20.8 [20.8(5)] | 0.5 [<1.4] |
| 8 | $^5D_3$ resonance at 2.58 MeV – No.4. | E2 | $^5S_2$ | 7 | 654.0477 | 1.1 | 2580 [2580(20)] | 55 [55(20)] |

The $P^2$ value determines coefficient in capture cross sections [1]. The number of resonance from the list given above, its energy and width with the given potentials is given for the initial state. The experimental values of energies and width of these resonances [5] are given in square brackets.

Table 2. The list of possible transitions from the $\{^{(2S+1)}L_J\}_i$ state to the different WF components $\{^{(2S+1)}L_J\}_f$ 3rd ES $^3S_1$ of $^{12}$B at the neutron capture on $^{11}$B and Gaussian parameters for initial scattering states.

| No. | $\{^{(2S+1)}L_J\}_i$ | Transition | $\{^{(2S+1)}L_J\}_f$ | $P^2$ | $V_0$, MeV | $\alpha$, fm$^{-2}$ | $E_r$, keV | $\Gamma_r$, keV |
|---|---|---|---|---|---|---|---|---|
| 1 | $^3P_0$ nonresonance scattering wave. | E1 | $^3S_1$ | 1 | 147.1709 | 0.18 | --- | --- |
| 2 | $^3P_1$ nonresonance scattering wave. | E1 | $^3S_1$ | 3 | 194.68751 | 0.22 | --- | --- |
| 3 | $^3P_2$ resonance at 430 keV – No.2. | E1 | $^3S_1$ | 5 | 11806.017 | 15.0 | 430 [430(10)] | 37 [37(5)] |
| 4 | $^3D_1$ resonance at 1.027 MeV – No.3. | E2 | $^3S_1$ | 3 | 1611.95103 | 1.25 | 1027 [1027(11)] | 9 [9(4)] |
| 5 | $^3D_1$ resonance at 4.8 MeV – No.6. | E2 | $^3S_1$ | 3 | 4502.245 | 3.5 | 4.80 [4.8] | 86 [90] |
| 6 | $^3D_2$ resonance at 4.7 MeV – No.5. | E2 | $^3S_1$ | 5 | 6444.382 | 5.0 | 4.70 [4.7] | 49 [45] |
| 7 | $^3D_3$ resonance at 20.8 keV – No.1. | E2 | $^3S_1$ | 7 | 16.0578 | 0.0125 | 20.8 [20.8(5)] | 0.5 [<1.4] |
| 8 | $^3D_3$ resonance at 2.58 MeV – No.4. | E2 | $^3S_1$ | 7 | 654.0477 | 1.1 | 2580 [2580(20)] | 55 [55(20)] |

The $P^2$ value determines coefficient in capture cross sections [1]. The number of resonance from the list given above, its energy and width with the given potentials is given for the initial state. The experimental values of energies and width of these resonances [5] are given in square brackets.

Table 3. The list of possible transitions from the $\{^{(2S+1)}L_J\}_i$ state to the different WF components $\{^{(2S+1)}L_J\}_f$ 4$^{th}$ ES $^3P_0$ of $^{12}$B at the neutron capture on $^{11}$B and Gaussian parameters for initial scattering states.

| No. | $\{^{(2S+1)}L_J\}_i$ | Transition | $\{^{(2S+1)}L_J\}_f$ | $P^2$ | $V_0$, MeV | α, fm$^{-2}$ | $E_r$, keV | $\Gamma_r$, keV |
|---|---|---|---|---|---|---|---|---|
| 1 | $^3S_1$ nonresonance scattering wave | E1 | $^3P_0$ | 1 | 5.61427 | 0.04 | --- | --- |
| 2 | $^3P_1$ nonresonance scattering wave | M1 | $^3P_0$ | 2 | 194.68751 | 0.22 | --- | --- |
| 3 | $^3P_2$ resonance at 430 keV – No.2. | E2 | $^3P_0$ | 2 | 11806.017 | 15.0 | 430 [430(10)] | 37 [37(5)] |
| 4 | $^3D_1$ resonance at 1.027 MeV – No.3. | E1 | $^3P_0$ | 2 | 1611.95103 | 1.25 | 1027 [1027(11)] | 9 [9(4)] |
| 5 | $^3D_1$ resonance at 4.8 MeV – No.6. | E1 | $^3P_0$ | 2 | 4502.245 | 3.5 | 4.80 [4.8] | 86 [90] |

The $P^2$ value determines coefficient in capture cross sections [1]. The number of resonance from the list given above, its energy and width with the given potentials is given for the initial state. The experimental values of energies and width of these resonances [5] are given in square brackets.

## 4. $n^{11}$B interaction potentials

For all partial potentials, that is, interactions for each orbital momentum $L$ at the given $J$ and $S$, the Gaussian form is used

$$V(^{2S+1}L_J, r) = -V_0(^{2S+1}L_J)\exp\{-\alpha[^{2S+1}L_J)r^2\}, \qquad (4)$$

where the depth $V_0$ and the width α of the potential depend on the momenta $^{2S+1}L_J$ of each partial wave. In some cases, this potential may also depend explicitly on the Young diagrams {*f*} and be different in the discrete and continuous spectrum, since in such states these diagrams are different [22].

Practically, all these potentials were already considered in our works [1,2], but here we briefly repeat basic principles used for their construction. In particular, the values for spectroscopic factor $S_f = 0.33$ and ANC $A_{NC} = 1.28(6)$ fm$^{-1/2}$ are given in [24] for the 2$^{nd}$ ES $^5S_2$, then using expressions (1)–(3) we have $C_w = 3.0(1)$ at $\sqrt{2k_0} = 0.742$. In that case, for parameters of this potential one can obtain values [1,2] listed in Table 1 under No. 1. This potential leads to the binding energy of -1.6964 MeV, gives AC equals 3.1(1), the mass radius of 2.78 fm and the charge radius of 2.45 fm, its phase shift is shown in Fig. 1 by the green solid curve.

The next values: $S_f = 0.63$ and $A_{NC} = 1.05(5)$ fm$^{-1/2}$ are given in work [24] for the 3$^{rd}$ ES $^3S_1$, then we have $C_w = 2.2(1)$ at $\sqrt{2k_0} = 0.605$. For parameters of the $^3S_1$ potential without FS the values given in Table 3 under No. 1 can be obtained. The binding energy of -0.7492 MeV absolutely coincided with the experimental value [5] was obtained for this potential, the charge radius of 2.47 fm, the mass radius of 2.94 fm and dimensionless AC equals 1.9(1) were obtained too. The shape of the $^3S_1$ scattering phase shift with this potential is shown in Fig. 1a by the blue solid line.

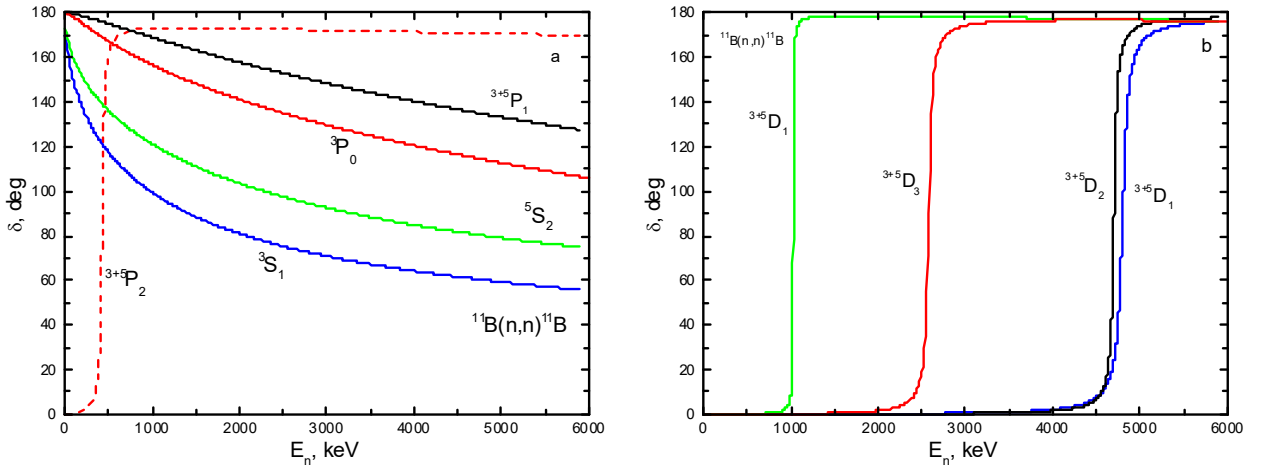

Fig. 1. Phase shifts of the elastic $n^{11}$B scattering at low energies obtained using potentials with parameters listed in Tables 1, 2 and 3.

The values $S_f = 0.113$ and $A_{NC} = 0.15(1)$ fm$^{-1/2}$ are given in work [24] for the 4$^{th}$ ES $^3P_0$, then we have $C_w = -0.45(1)$ and also negative $C_w = -0.76(1)$ at $\sqrt{2k_0} = 0.583$, because this state contains FS. It is possible to obtain values given in Table 2 under No. 1 for parameters of the $^3P_0$ potential with FS. The binding energy of -0.6470 MeV, absolutely coincided with the experimental value [5], the charge radius of 2.45 fm, the mass radius of 2.75 fm and dimensionless AC equals 0.75(1) are obtained with this potential. The shape of the scattering phase shift with this potential is shown in Fig. 1a by the red solid curve. These ES potentials will use not only for description of the BSs themselves, but for description of the nonresonance $n^{11}$B scattering in these partial waves.

The GS potential of $^{12}$B in the $n^{11}$B channel [1] is used for the nonresonance $^{3+5}P_1$ scattering wave. New data are given for it in [24], that is $S_f = 0.69$ and $A_{NC} = 1.15(6)$ fm$^{-1/2}$. Therefore, we obtain for the range of the dimensioned AC 1.31–1.38 fm$^{-1/2}$, which gives for the dimensionless AC $C_w$ of the form (3) the range of 1.49–1.57 at the average value of 1.53(4). Let us construct now the GS potential with FS, which corresponds to the AC from this range – its parameters are listed in Table 1 under No. 2. This potential gives the negative AC of -1.64(1), the charge radius of 2.43 fm and the mass radius of 2.53 fm at the binding energy of -3.37000 MeV with an accuracy of $10^{-5}$ MeV [3,4], which absolutely coincides with the experimental value [5]. The phase shift of such potential is shown in Fig. 1a by the black solid curve. The matter radius of 2.41(3) fm, obtained in [25], can be used for comparison of radii.

For the potential of the resonance $^{3+5}P_2$ wave with one bound FS, parameters No. 3 in Tables 1, 2 and 3 were obtained. This potential leads to a resonance energy $E_n = 430(1)$ keV with a width of 37(1) keV (c.m.), which completely coincide with the experimental data [5]. The scattering phase shift turned out to be 90(1) for this energy. The shape of the resonance mixed $^{3+5}P_2$ scattering phase shift is shown in Fig. 1a by the red dashed curve.

Here we must remember that if the potential contains $N + M$ forbidden and allowed states, it obeys the Levinson generalized theorem and its phase shift at zero energy begins with $\pi \cdot (N + M)$ [23]. However, in Fig. 1a, the $P_1$ phase shift, having the bound FS and the bound AS, is from 180°, not from 360°, and the $P_2$ phase shift is from 0 degrees, not from 180° for a more familiar presentation of the results and placement of all phase shifts in one figure.

For $^{3+5}D_2$, $^{3+5}D_1$ resonances at 4.7 and 4.8 MeV, respectively, potentials with the FSs No.6 and No.5 were obtained. The first of them gives a width of 49(1) keV, and the second one is of 86(1) keV, which agrees well with the data [5], and energies exactly coincide with those given above. The phase shifts of these potentials are shown in Fig. 1b by the black and blue solid curves – at resonance they have a value of 90(1)°.

Resonance at 1.027 MeV was also able to be reproduced only under the assumption of the $^{3+5}D_1$ wave and parameters with the FS No. 4 in Table 1, which lead to a width of 9(1) keV with exactly coincident energy and scattering phase shift, which is shown in Fig. 1b by the green solid curve, which in resonance is also equal to 90(1)°.

Resonance at 2.58 MeV in the $^{3+5}D_3$ wave leads to the potential listed in Table 1 under No. 8 at exactly coincident energy and the width of 55(1) keV. The phase shift of such potential is shown in Fig. 1b by the red solid curve.

Potential parameters for the 1$^{st}$ RS in the $^{3+5}D_3$ wave at 20.8 keV are given, for example, in Table 1 under No. 7. They lead to the accurate resonance energy the width of 0.5(1) keV in c.m. – note that the accurate width of this resonance is not known.

## 5. Total cross section of the neutron capture on $^{11}$B to the 2$^{nd}$ ES and reaction rate

The calculation results of the total cross sections for all transitions from Table 1 with the listed in it scattering potentials, using for the 2$^{nd}$ ES potential No. 1, are shown in Fig. 2 by the black solid curve. The cross sections were calculated to neutron energy up to 7 MeV, but, as it was said, do not take into account resonances above 5 MeV. As it was mentioned above, the $M$1 transition No. 1 from Table 1 leads to zero cross sections, because one and the same potential uses for it in continuous and discrete spectra. The red dashed curve shows the calculation result of the $E$1 capture cross sections from the resonance $^{3+5}P_2$ wave No. 3 from Table 1. The value of this cross section in the maximum of the first resonance is at the level of 113 μb. The blue dashed curve shows $E$2 cross sections for resonance at 20.8 keV No. 7 from Table 1, the value of which reaches only 0.11 μb at 3.2 μb for summarized cross sections. This transition practically does not influence on summarized cross sections of this process, as well as the resonance at 1027 keV No. 4, cross section of which is lower than $10^{-2}$ μb. Contributions of resonances Nos. 5, 6 and 8 from Table 1 also do not give visible contribution to the total cross sections and cannot be seen in Fig. 2.

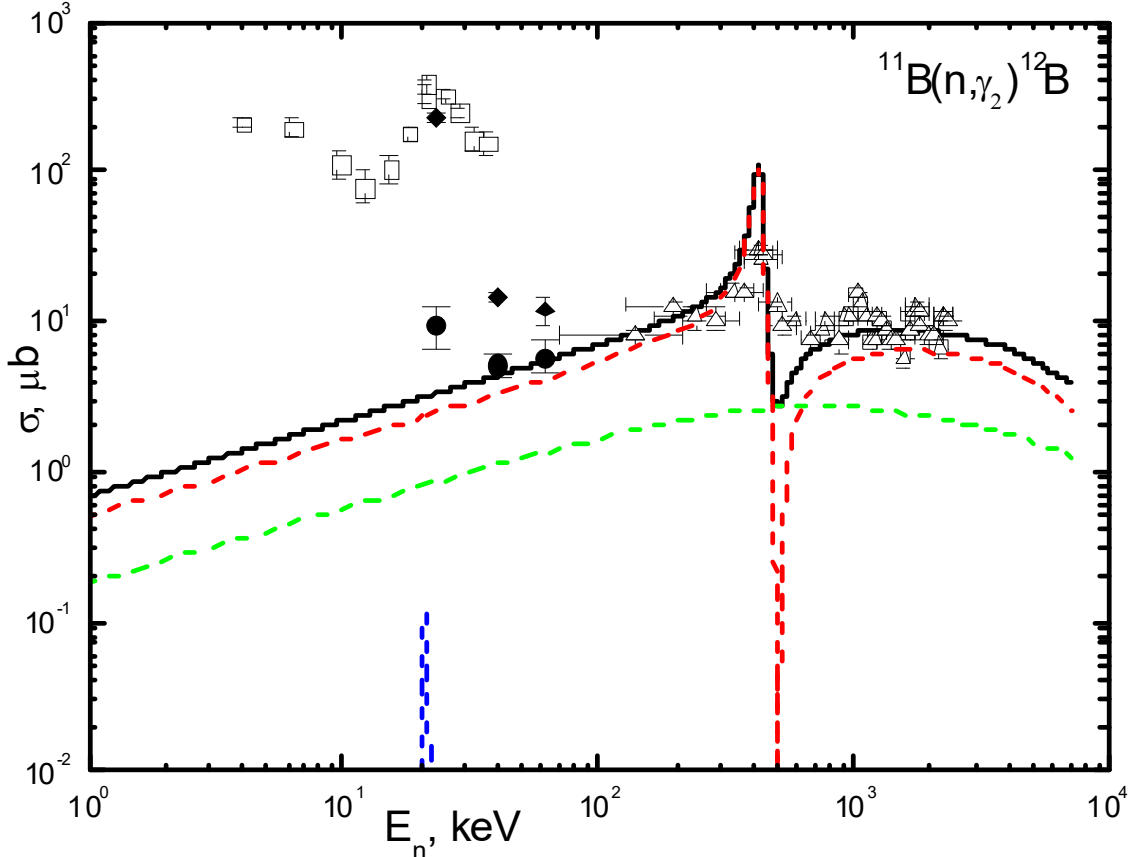

Fig. 2. Total cross sections of the neutron radiative capture reaction on $^{11}$B for transition to the 2$^{nd}$ ES in the energy range of 1 keV to 7 MeV. Experimental data: points (●) – total cross sections to the GS from [26], rhombs (♦) – total cross sections to the 2$^{nd}$ ES from [26], open triangles (Δ) – total cross sections from [9], open squares (□) – total capture cross sections from [27]. Curves – calculation results for different transitions from Table 1 with listed in it potentials of the initial channel.

The available experimental data in the range of 23 meV to 60 keV are shown in Fig. 2: points – for capture to GS and rhombs – to the 2$^{nd}$ ES [26]. Open triangles – total capture cross sections from [9], open squares – total capture cross sections from [27]. As seen from Fig. 2, the resonance at 430 keV and cross sections at higher energies is described correctly in general. However, data from [26] for the 2$^{nd}$ ES, shown in Fig. 2 by rhombs, lie in absolutely other range for values of total cross sections.

Furthermore, the reaction rate of the neutron capture on $^{11}$B was calculated, and in units of cm$^3$ mol$^{-1}$s$^{-1}$ it can be represented as [16]

$$N_A \langle \sigma v \rangle = 3.7313 \cdot 10^4 \mu^{-1/2} T_9^{-3/2} \int_0^\infty \sigma(E) E \exp(-11.605 E / T_9) dE \quad , \qquad (5)$$

where $E$ is given in MeV, the total cross section σ($E$) is measured in μb, μ is the reduced mass given in amu, and $T_9$ temperature is given in $10^9$ K [16]. To obtain this integral, the calculated total cross section for 7000 points in the energy range of $10^{-5}$ to $7 \cdot 10^3$ keV was used.

Results of the reaction rate of 0.01 to 10 $T_9$ are shown in Fig. 3 by the black solid curve. As seen from Fig. 3 accounting of the second resonance at 430 keV visibly influence to shape of the reaction rate, leads it to the resonance form. For comparison, the results from [28] are shown by the blue dotted-dashed curve. Calculation of this rate is carried out taking into account resonances and capture to the ESs. The green dashed curve shows obtained by us reaction rate for the neutron capture on $^{11}$B to the GS of $^{12}$B [1].

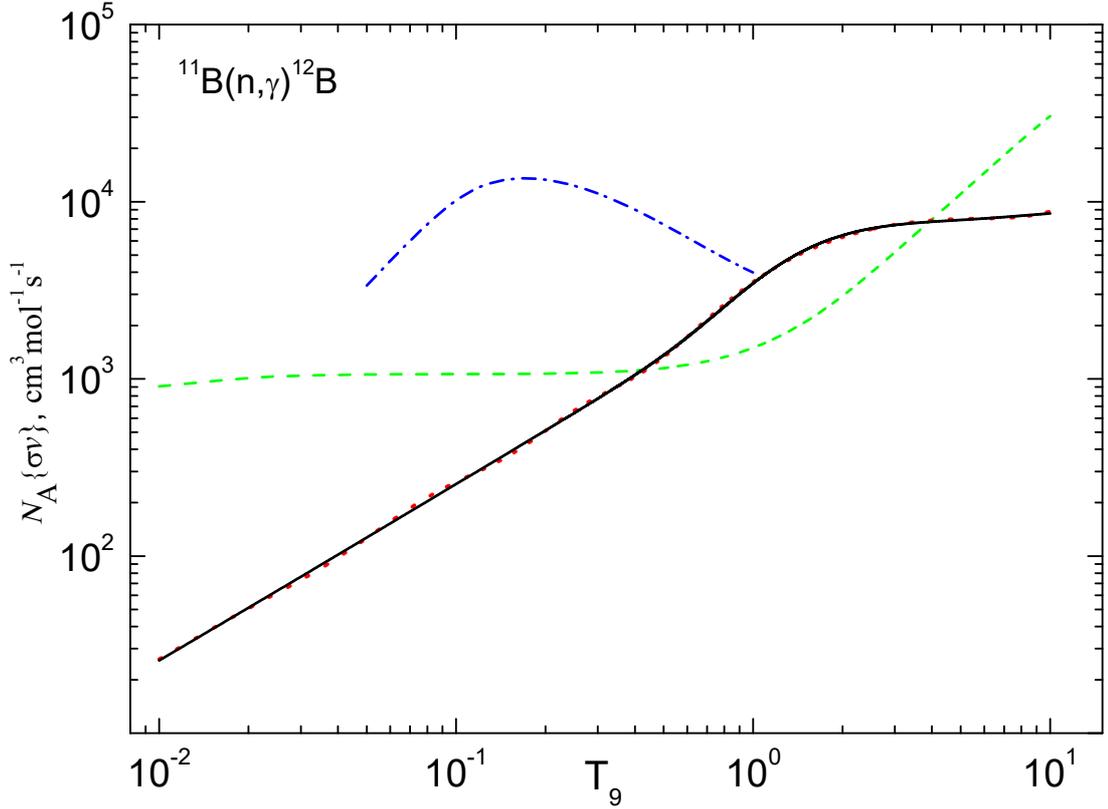

Fig. 3. Reaction rate of the radiative neutron capture on $^{11}$B to the 2$^{nd}$ ES. The designations of the curves are given in the text.

Furthermore, the parameterization of the calculated reaction rate, which is shown in Fig. 3 by the black solid curve, was carried out by the expression of the form [29]

$$N_A \langle \sigma v \rangle = a_1 / T_9^{2/3} \cdot \exp(-a_2 / T_9^{1/3}) \cdot (1.0 + a_3 \cdot T_9^{1/3} + a_4 \cdot T_9^{2/3} + a_5 \cdot T_9 +$$
$$+ a_6 \cdot T_9^{4/3} + a_7 \cdot T_9^{5/3} + a_8 \cdot T_9^{7/3}) + a_9 / T_9^{1/2} \cdot \exp(-a_{10} / T_9^{1/2}) +$$
$$+ a_{11} / T_9 \cdot \exp(-a_{12} / T_9) + a_{13} / T_9^{3/2} \cdot \exp(-a_{14} / T_9^{3/2}) \quad (6)$$

with parameters listed in Table 4.

Table 4. Parameters of analytical parametrization of the reaction rate.

| No.   | 1       | 2       | 3        | 4        | 5         | 6         | 7         |
|-------|---------|---------|----------|----------|-----------|-----------|-----------|
| $a_i$ | 2.37006 | 3.20385 | 49224.07 | 17619.94 | -11246.65 | -3055.331 | -25.58309 |
| No.   | 8       | 9       | 10       | 11       | 12        | 13        | 14        |
| $a_i$ | 321.4723| 66.72196| 0.32461  | -981.7858| 0.46404   | -1946.273 | 0.66658   |

The χ² value turns out to be equal to 0.09 at 5% errors of the calculated reaction rate, and approximate curve of parametrization practically coincides with the black solid curve, as was shown in Fig. 3 by the red dotted curve.

## 6. Total cross section of the neutron capture on $^{11}$B to the 3$^{rd}$ ES and reaction rate

The calculation results of the total cross sections for all transitions from Table 2 with the listed in it scattering potentials and BSs are shown in Fig. 4 by the black solid curve. The cross sections were calculated to neutron energy up to 7 MeV, but do not take into account resonances above 5 MeV. The red dashed curve shows in Fig. 4 the $E$1 transitions under Nos. 1,2,3 from Table 2. The nonresonance $E$1 transitions under Nos. 1,2 from Table 2 are shown by the green dashed curve in Fig. 4. Calculation results for cross sections of the $E$2 capture from $D$ waves No. 4–No.8 from Table 2 cannot be seen in Fig. 4. It is seen, from the given results that only one resonance at 430 keV is clearly observed, its value is at the level of 26 µb. Contributions of two other nonresonance waves, shown in Fig. 4 by the green dashed curve are relatively small, and the $E$2 cross sections have negligible values.

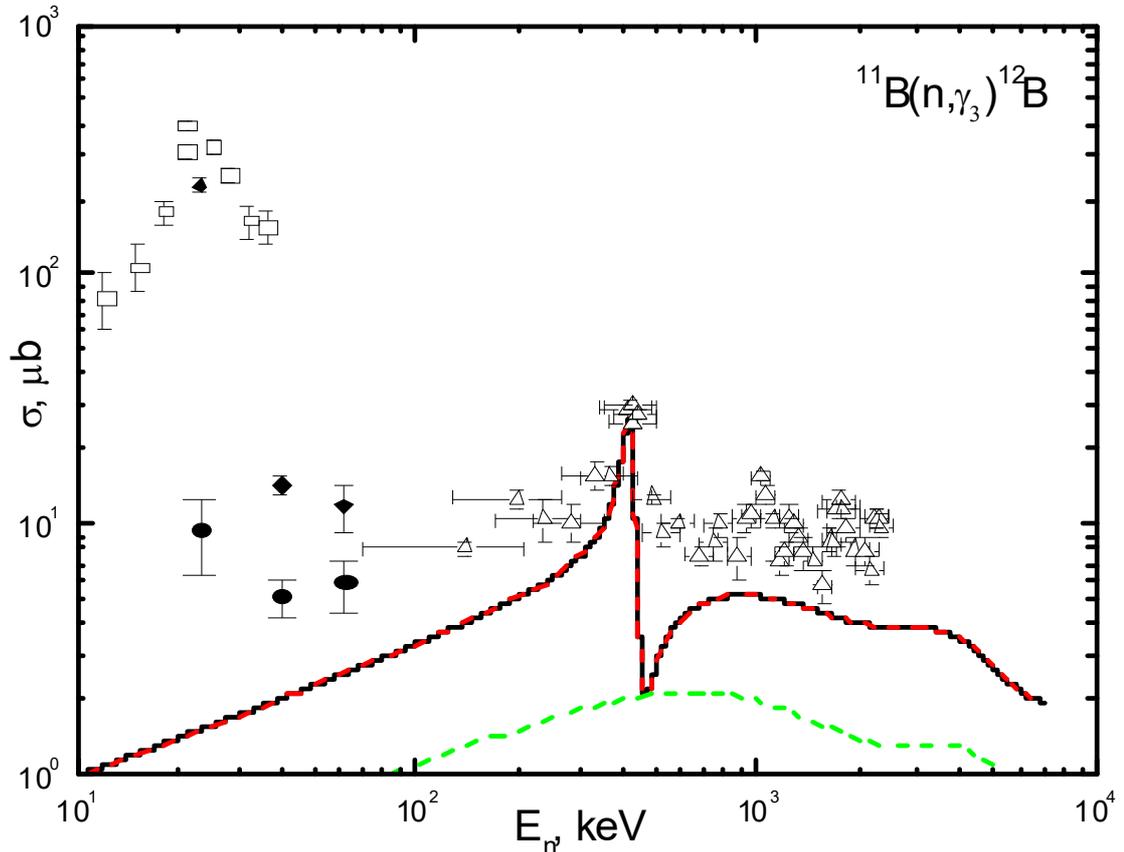

Fig. 4. Total cross sections of the neutron radiative capture reaction on $^{11}$B for transition to the 3$^{rd}$ ES in the energy range of 10 keV to 7 MeV. Experimental data are described in Fig. 2. Curves – calculation results for different transitions from Table 2 with listed in it potentials of the initial channel.

Furthermore, the reaction rate of the neutron capture on $^{11}$B was calculated (5). To obtain this integral, as previously, the calculated total cross section for 7000 points in the energy range of $10^{-5}$ to $7 \cdot 10^3$ keV was used. Results of the reaction rate are shown in Fig. 5 by the black solid curve. The rise of the reaction rate caused by the resonance at 430 keV is in a good light here, meanwhile the increase starting influence at temperatures above $1 \cdot T_9$. As seen from Figs. 4 and 5 accounting of this resonance noticeably influences to the total cross sections and affects to the shape of reaction rate. The results from [28] are shown by the blue dotted-dashed curve for comparison. The calculation of this reaction rate is carried out taking into account resonances and captures to the ESs. Green dashed curve shows the reaction rate of the neutron capture on $^{11}$B to the GS of $^{12}$B [1] obtained by us

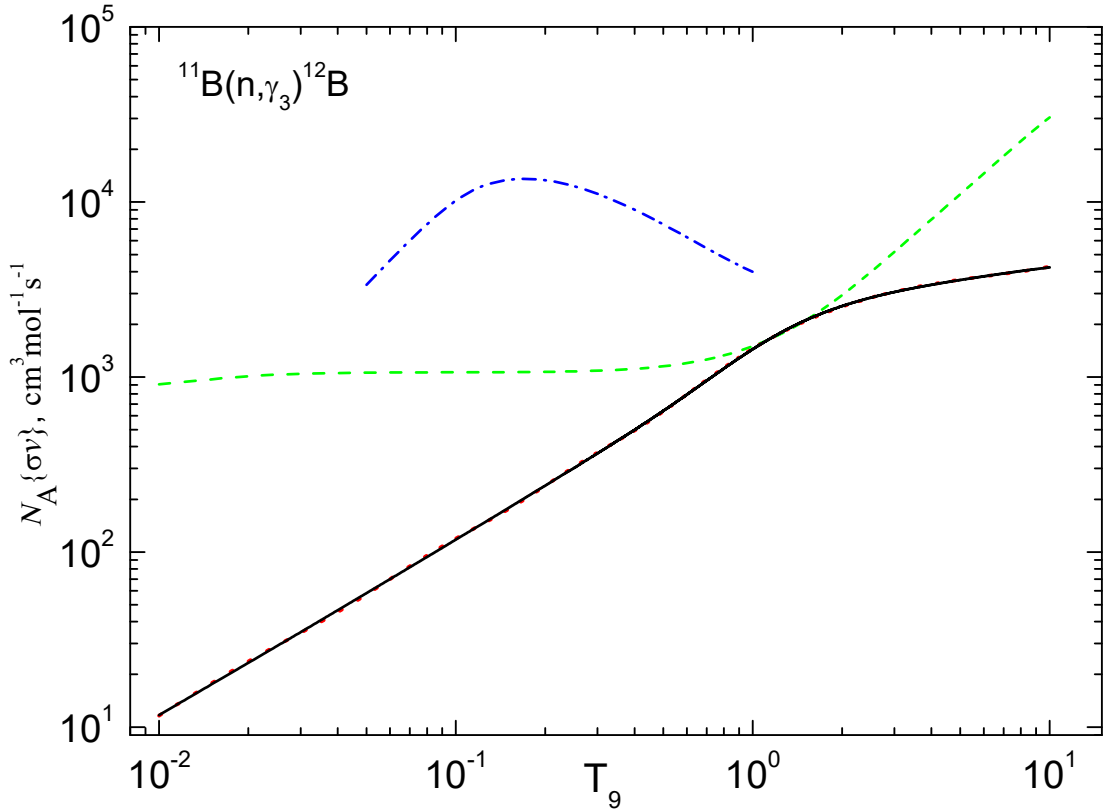

Fig. 5. Reaction rate of the radiative neutron capture on $^{11}$B to the 3$^{rd}$ ES. The designations of the curves are given in the text.

Furthermore, the parameterization of the calculated reaction rate, which is shown in Fig. 5 by the black solid curve, by the expression (6) with parameters listed in Table 5. The $\chi^2$ value turns out to be less than 0.02 at 5% errors of the calculated reaction rate, and approximate curve, shown by the red dotted curve, practically coincides with the solid curve.

Table 5. Parameters of analytical parametrization of the reaction rate.

| No. | 1 | 2 | 3 | 4 | 5 | 6 | 7 |
|---|---|---|---|---|---|---|---|
| $a_i$ | 0.71385 | 3.11604 | 46682.13 | 21767.36 | -9057.345 | -1769.751 | 164.3489 |
| No. | 8 | 9 | 10 | 11 | 12 | 13 | 14 |
| $a_i$ | 178.282 | 37.85767 | 0.34924 | -294.4681 | 0.48795 | -446.9756 | 0.66007 |

## 7. Total cross section of the neutron capture on $^{11}$B to the 4$^{th}$ ES and reaction rate

The calculation results of the total cross sections for all transitions from Table 3 with the listed in it scattering potentials and BSs are shown in Fig. 6 by the black solid curve. The calculation result for cross sections of the $E1$ capture from $D$ waves No. 4, No.5 from Table 3 is shown by the green dashed curve. Two resonances at 1027 keV and 4.8 MeV can be seen from these results, the value of the first is on the level of 10 μb the second is about 3 μb. The blue dashed curve shows results for transition No. 1 from Table 3. Results of cross sections for sum of transitions No. 2 and No. 3 from Table 3 are shown by the red dashed curve. The black dashed curve shows results for transition No. 1 and No. 2 from Table 3. They show a small contribution of the $M1$ process No. 2 to the nonresonance part of total cross sections.

Furthermore, the reaction rate of the neutron capture on $^{11}$B (5) was calculated. Results for the reaction rate are shown in Fig. 7 by the black solid curve. Here, the rise of reaction rate, caused by the resonance at 1027 keV, is well observed, meanwhile increase starting influence at temperatures more than $1 \cdot T_9$. As seen from Figs. 6 and 7, the accounting of this resonance noticeably influence to the total cross sections and to the reaction rate shape. The calculation of this rate is carried out taking into account resonances and capture to the ESs. The green dashed curve shows obtained by us reaction rate of the neutron capture on $^{11}$B to the GS of $^{12}$B [1].

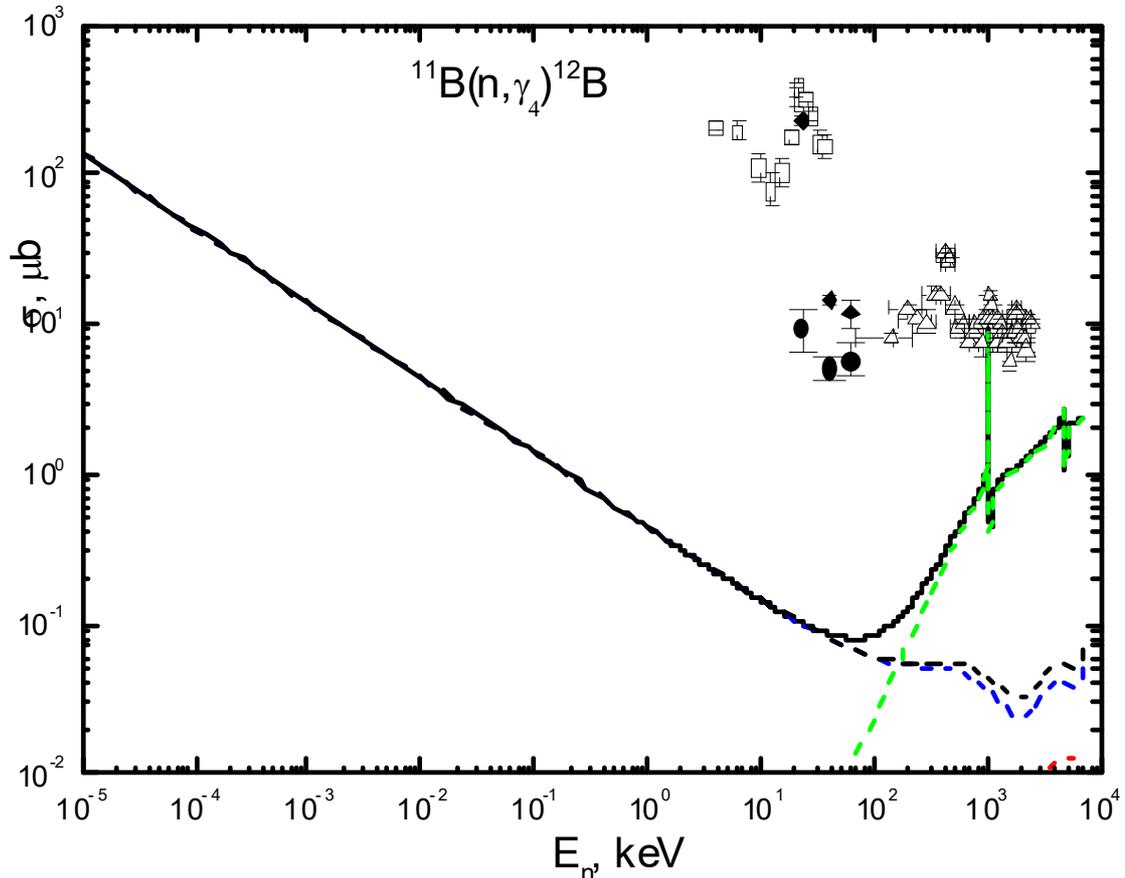

Fig. 6. Total cross sections of the neutron radiative capture reaction on $^{11}$B for transition to the 4$^{th}$ ES in the energy range of 10 meV to 7 MeV. Experimental data are described in Fig. 2. Curves – calculation results for different transitions from Table 3 with listed in it potentials of the initial channel.

Furthermore, the parameterization of the calculated reaction rate, which is shown in Fig. 7 by the black solid curve, by the expression (6) with parameters listed in Table 6.

Table 6. Parameters of analytical parametrization of the reaction rate.

| No. | 1 | 2 | 3 | 4 | 5 | 6 | 7 |
|---|---|---|---|---|---|---|---|
| $a_i$ | 0.02831 | 2.95917 | -607.2552 | 23045.51 | 11493.88 | 10095.68 | 5682.402 |
| No. | 8 | 9 | 10 | 11 | 12 | 13 | 14 |
| $a_i$ | -112.9582 | 5.82467 | 0.18256 | -82.21239 | 1.23343 | -228.3177 | 2.96388 |

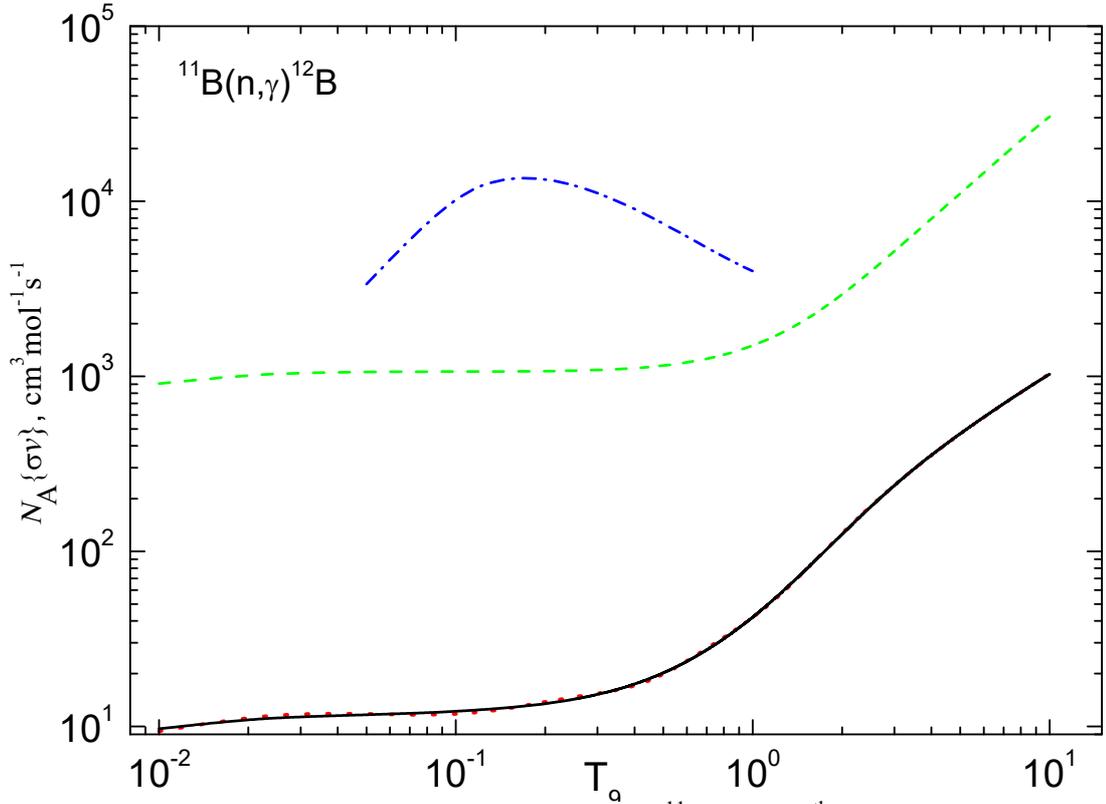

Fig. 7. Reaction rate of the radiative neutron capture on $^{11}$B to the 4$^{th}$ ES. The designations of the curves are given in the text.

The $\chi^2$ value turns out to be equal to 0.02 at 5% errors of the calculated reaction rate, and approximate red dotted curve practically coincides with the theoretical solid curve.

## 8. Total cross sections and reaction rates taking into account all ESs and resonances

Reproduce now summarized cross sections for capture to all ESs, obtained here and in [1,2]. These results are shown in Fig. 8a by the blue solid curve. Red solid curve shows cross sections for capture to the GS for the AC value of 1.02 [1]. Green dashed curve shows cross sections for capture to the 1$^{st}$ ES at AC = 1.24 and the width of the first resonance of 0.5 keV [2]. Blue dashed curve shows cross sections for capture to the 2$^{nd}$ ES from Fig. 2. Red dashed curve shows cross sections for capture to the 3$^{rd}$ ES from Fig. 4. Violet dashed curve shows cross sections for capture to the 4$^{th}$ ES from Fig. 6. It is seen from these results that in the energy range above 100 keV calculated cross sections lie higher of the available experimental data [9] in 2–3 times. It is connected with the fact that the capture to the 2$^{nd}$ ES already correctly describes these data and the capture to other ESs and GS strongly increase these cross sections. Exactly capture to the GS [1] gives the sharp rise of cross sections at energies more than 1 MeV, shown in Fig. 8a by the red solid curve. However, there is a good agreement with data [27] at energies in the energy range of 12–36 keV, although first three experimental points lead to the higher cross sections than the theoretical calculation.

If to use nonresonance partial waves from Table 1, 2 and 3 in calculations, then the results shown in Fig. 8b are obtained – designations of curves and data as in Fig. 8a. It is seen that summarized cross section shown by the blue solid curve goes at the lowest part of data [9]. These results are, apparently, analogue of result of [30], where the theoretical nonresonance part of cross sections was obtained, which also goes at the lowest points of data [9].

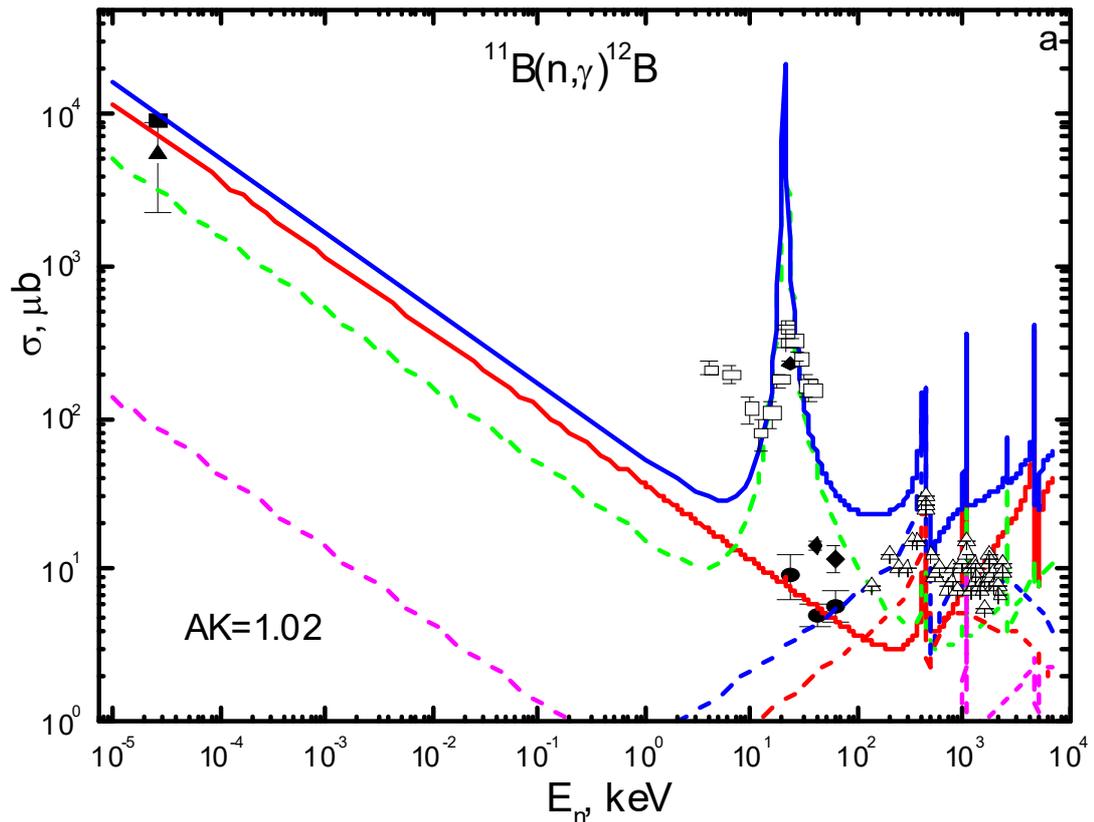

Fig. 8a. Total cross sections of the neutron radiative capture reaction on $^{11}$B for transitions to all ESs in the energy range of 10 meV to 7 MeV. Experimental data are described in Fig. 2. Curves – calculation results for different transitions from Tables 1, 2, 3 and works [1,2] with listed in them potentials of the initial channel.

Let us consider now possible causes of disarrangement in cross sections [9,27] with our results shown in Figs. 8a,b:

1) data errors of [27] at energies lower 12 keV (Fig. 8a) have, apparently, larger value than the value that given in this work, because experimental points demonstrate rise of the cross sections at energies lower than 12 keV, which similar to the resonance, but there is no such resonance in modern spectra of $^{12}$B [5];

2) one of the possible causes of exceedance the cross sections over data [9] – incompletely correct AC values, as a minimum for the 2$^{nd}$ ES, and possibly for the GS – however calculated cross sections weakly depend on them;

3) old and not quite accurate data of experimental measurements of the total cross sections in [9] do not take into account capture to all ESs. In particular, in this work there was no direct measurement of γ-quantum spectra – only β decay $^{12}$B(β) → $^{12}$C were registered. Besides, it was noted in work [9] that the comparison of the radiative widths, obtained in the experiment, and their theoretical estimation shows disarrangement of one to two orders – theory gives higher values;

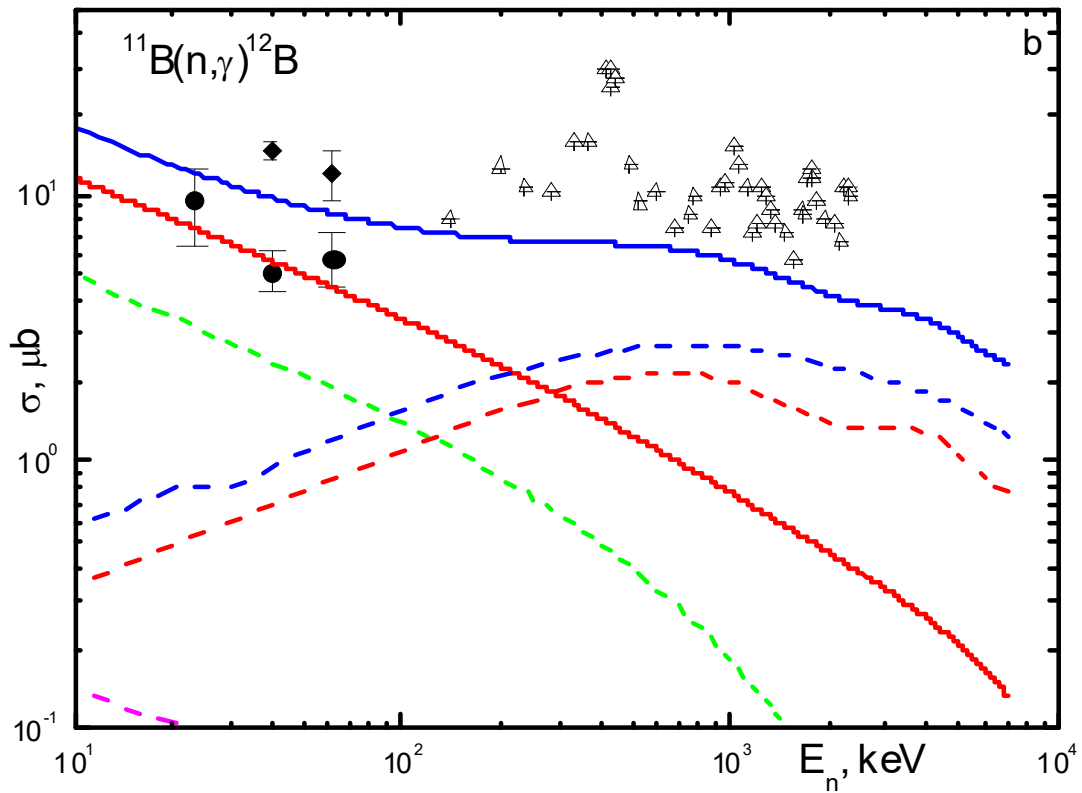

Fig. 8b. Total cross sections of the radiative neutron capture on $^{11}$B for transitions to all ESs in the energy range of 1 keV to 7 MeV. Experimental data are described in Fig. 2. Curves – calculation results for different transitions from Tables 1, 2, 3 and works [1,2] with listed in them potentials of the initial channel.

    4) comparison of results [9] and calculations [31], as was noted in [9], also leads to the discrepancy of widths up to two times;
    5) possible inaccuracies and incorrect predictions of our model at energies above 2 MeV, because usually it is tested at energy range up to 1–2 MeV [3,4,12];
    6) besides, it should be noted that in [32] the calculations of the potential capture to the 2$^{nd}$ and 3$^{rd}$ ESs of $^{12}$B were carried out and already at energy 1 MeV calculation results rise to 25 μb, that is, exceed data [9] in 2–3 times, if to take for them 10 μb on the average.

    As it is seen, there are quite a lot causes for doubts in results of [9], however, as it was said above, other, more precise measurements for this energy range are absent in data bases [10] or [11]. Also there are no sufficiently detailed theoretical calculations taking into account all ESs and all possible resonances at low energies. At the same time, this work finalizes our considering of the neutron capture processes on $^{11}$B [1,2] and gives the estimation of cross sections taken into account all ESs and all possible resonances at energies up to 5 MeV. These results can be well stimulus for further experimental study of this capture reaction.
    Furthermore, as seen from Fig. 8a the calculated cross section taken into account

captures to all ESs practically is a straight line at energies of 10 meV to 1 keV and it can be approximated by a simple function of the next form

$$\sigma_{ap}(\mu b) = \frac{A}{\sqrt{E(\text{keV})}} \quad (7)$$

The value of constant $A = 53.251$ µb·keV$^{1/2}$ is determined according to one point in calculated cross sections (blue solid curve in Fig. 8a) at minimal energy equals 10 meV. The cross section value at thermal energy 25.3 meV is equal to 10.6 mb. It can be seen from here that the contribution of all other transitions to other ESs, except the capture to the 1$^{st}$ ES, at thermal energy approximately equal to 9.1(1) mb [15], that is, for the purpose of normalize the cross sections to the experimental value at 25.3 meV it is necessary to multiply the calculated cross section on 0.86. This coefficient can be used for all other energies, and also for the total reaction rate with the capture to the GS and to all ESs.

The module

$$M(E) = \left| [\sigma_{ap}(E) - \sigma_{theor}(E)] / \sigma_{theor}(E) \right| \quad (8)$$

of relative deviation of the calculated theoretical cross section ($\sigma_{theor}$) and approximation ($\sigma_{ap}$) of this cross section by the given above function in the range up to 1 keV does not exceed 0.01%.

Furthermore, the summarized reaction rate of the neutron capture on $^{11}$B shown in Fig. 9 by the black solid curve was calculated. It is obtained on the basis of work [2] for potential with AC = 1.24 and results for the capture to the GS for potential with AC = 1.02 [1], and also on all reaction rates obtained above in this paper. Red dashed curve shows the capture reaction rate to the 1$^{st}$ ES in Fig. 9 [2]. The capture reaction rate to the GS is shown by the green dashed curve [1]. Black dashed curve shows the capture reaction rate to the 2$^{nd}$ ES. The capture reaction rate to the 3$^{rd}$ ES is shown by the blue dashed curve. The capture reaction rate to the 4$^{th}$ ES is shown by the violet dashed curve. Blue dotted-dashed curve shows the capture reaction rate from [28]. The approximation of the total reaction rate by the form [29] was carried out.

$$N_A \langle \sigma v \rangle = a_1 / T_9^{2/3} \exp(-a_2 / T_9^{1/3})(1.0 + a_3 T_9^{1/3} + a_4 T_9^{2/3} + a_5 T_9 + a_6 T_9^{4/3} + a_7 T_9^{5/3} +$$
$$+ a_8 T_9^{7/3}) + a_9 / T_9^{1/2} \exp(-a_{10} / T_9^{1/2}) + a_{11} / T_9 \exp(-a_{12} / T_9) + a_{13} / T_9^{3/2} \exp(-a_{14} / T_9^{3/2}) +$$
$$+ a_{15} / T_9^2 \exp(-a_{16} / T_9^2) + a_{17} T_9^{a_{18}} \quad (9)$$

with parameters from Table 7. The $\chi^2$ value turns out to be equal to 0.01 at 5% errors of the calculated reaction rate, and approximate green dotted curve practically coincides with the black theoretical solid curve.

Table 7. Parameters of analytical parametrization of the reaction rate.

| No.   | 1        | 2       | 3        | 4         | 5        | 6        | 7         | 8       | 9         |
|-------|----------|---------|----------|-----------|----------|----------|-----------|---------|-----------|
| $a_i$ | 53.90521 | 2.4133  | 34583.2  | -19082.22 | 661.9457 | 3466.372 | -939.6439 | 38.5252 | -366636.7 |
| No.   | 10       | 11      | 12       | 13        | 14       | 15       | 16        | 17      | 18        |
| $a_i$ | 1.40674  | 31401.31| 0.20459  | -15909.56 | 0.19121  | 4341.591 | 0.10113   | 75.3331 | -0.61643  |

It is seen from Fig. 9 that the reaction rate obtained by us, even taking into account coefficient of 0.86, is visibly higher the reaction rate from [13] or [28], shown in Fig. 9 by the blue dotted-dashed curve. For example, at temperature 0.1 $T_9$ it exceeds results [28] in 3.7 times, and taking into account 0.86 in 3.2 times. At temperature 1 $T_9$ it is higher in 4.2 times, and taking into account coefficient in 3.6 times. In the maximum at temperature of 0.15 $T_9$ the curve from [28] is less in 3.5 times, and taking into account coefficient in 3.0 times.

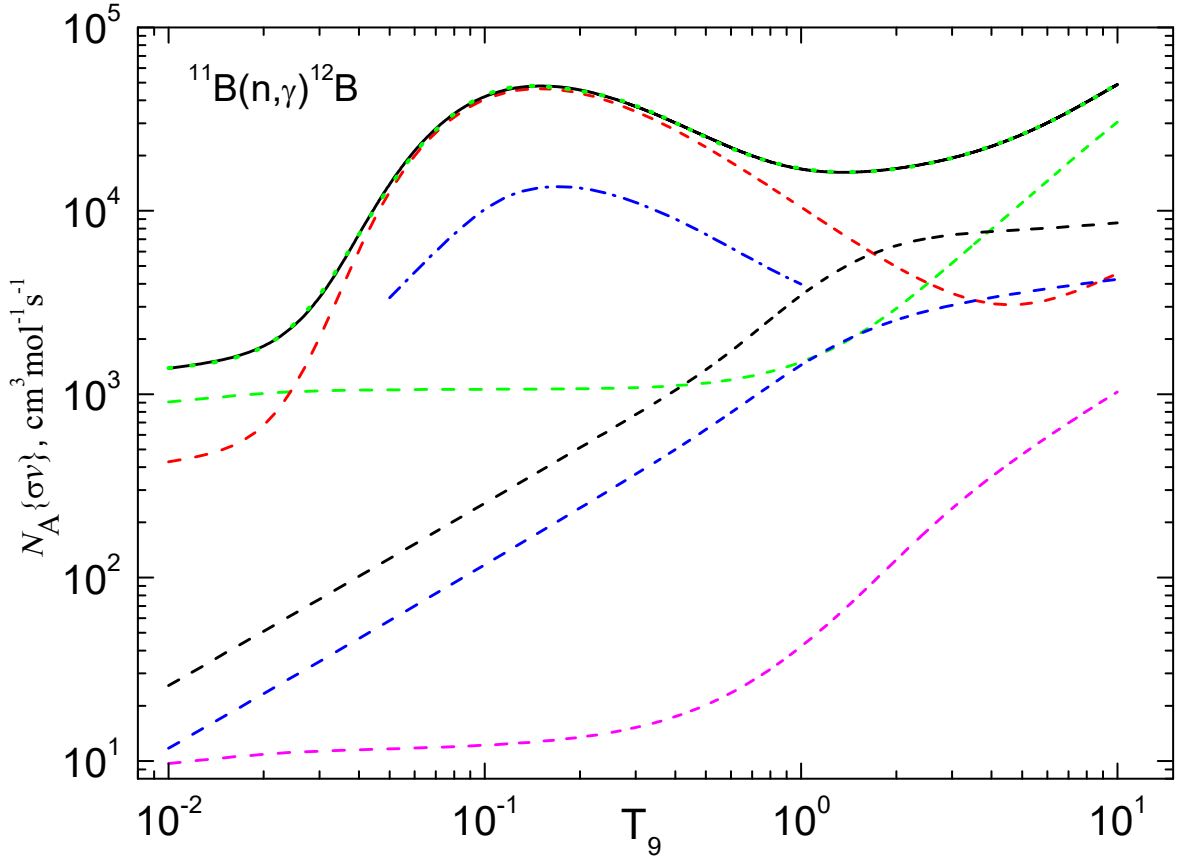

Fig. 9. Reaction rate of the radiative neutron capture on $^{11}$B to all ESs. The designations of the curves are given in the text.

## 9. Conclusion

The used potentials of the GS and all BSs of $^{12}$B in the n$^{11}$B channel preliminary are coordinated with their characteristics, including asymptotic constant and binding energy. The interactions of the second and third excited, but bound in the n$^{11}$B channel, states of $^{12}$B are used. Taking into account of the resonances up to 5 MeV increases total cross sections and reaction rate at temperature more than 0.02–0.03 $T_9$ starting rise and has clearly resonance shape, which guarantees it accounting of the first resonance at 20.8 keV for capture to the 1$^{st}$ ES [2].

New AC data given in [24] allow one to construct quite unambiguous potentials of the n$^{11}$B interaction [1,2]. Accounting of the first resonance influence to the reaction rate at the capture to the 1$^{st}$ ES very hard and, as a consequence, to the total summarized reaction rate of the neutron capture on $^{11}$B. It is noticeably differ from available earlier results for this reaction [13,28]. Our results can lead to the overestimation of the contribution of this reaction in the general balance of element formation in the BBN.

The obtained calculated cross sections taking into account resonances can be considered as a prediction of results for total cross sections and as a motive for new experimental studies of this reaction. New measurements and these results can noticeably influence to the efficiency results of $^{12}$B in the preliminary nucleosynthesis at the specific temperatures.

**Acknowledgements**

This work was supported by the Grant of Ministry of Education and Science of the Republic of Kazakhstan through the program BR05236322 "Study reactions of thermonuclear processes in extragalactic and galactic objects and their subsystems" in the frame of theme "Study of thermonuclear processes in stars and primordial nucleosynthesis" through the Fesenkov Astrophysical Institute of the National Center for Space Research and Technology of the Ministry of Defence and Aerospace Industry of the Republic of Kazakhstan (RK).